\documentclass[sigconf]{acmart}

\renewcommand\footnotetextcopyrightpermission[1]{} 
\usepackage{amsmath} 
\usepackage{amssymb}
\usepackage{amsfonts}
\usepackage{algorithm}
\usepackage{algpseudocode}
\usepackage{subcaption}
\usepackage{epstopdf}
\usepackage{graphicx}
\usepackage{booktabs} 
\usepackage{amsopn}
\usepackage{comment}
\usepackage{enumitem}
\usepackage{makebox}
\usepackage{etoolbox}
\usepackage{setspace}

\newcommand{\superscript}[1]{\ensuremath{^{\textrm{#1}}}}
\def\sharedaffiliation{
    \end{tabular}
    \begin{tabular}{c}}
\def\I{\superscript{1}}
\def\II{\superscript{2}}
\def\III{\superscript{3}}
\def\IV{\superscript{4}}
\def\cma{\superscript{,}}
\setcopyright{none}

\begin{document}
\title{Online Interactive Collaborative Filtering Using Multi-Armed Bandit with Dependent Arms}

\author{Qing Wang, Chunqiu Zeng, Wubai Zhou, Tao Li}
\affiliation{%
  \institution{Florida International University}
  \city{Miami}
  \state{FL, USA}
}
\email{qwang028, czeng001, wzhou005, taoli@cs.fiu.edu}

\author{Larisa Shwartz}
\affiliation{%
  \institution{IBM T.J. Watson Research Center}
  \city{Yorktown Heights}
  \state{NY, USA}
}
\email{lshwart@us.ibm.com}
\author{Genady Ya. Grabarnik}
\affiliation{%
  \institution{Dept. Math \& Computer Science, \\ St. John's University}
  \city{Queens}
  \state{NY, USA}
}
\email{grabarng@stjohns.edu}

\begin{abstract}
Online interactive recommender systems strive to promptly suggest to consumers appropriate items (e.g., movies, news articles) according to the current context including both the consumer and item content information. 
However, such context information is often unavailable in practice for the recommendation, where only the users' interaction data on items can be utilized.
Moreover, the lack of interaction records, especially for new users and items, worsens the performance of recommendation further. To address these issues, collaborative filtering (CF), one of the recommendation techniques relying on the interaction data only, as well as the online multi-armed bandit mechanisms, capable of achieving the balance between exploitation and exploration, are adopted in the online interactive recommendation settings, by assuming independent items (i.e., arms). Nonetheless, the assumption rarely holds in reality, since the real-world items tend to be correlated with each other (e.g., two articles with similar topics).

In this paper, we study online interactive collaborative filtering problems by considering the dependencies among items. We explicitly formulate the item dependencies as the clusters on arms, where the arms within a single cluster share the similar latent topics.
In light of the topic modeling techniques, we come up with a generative model to generate the items from their underlying topics. 
Furthermore, an efficient online algorithm based on particle learning is developed for inferring both latent parameters and states of our model. 
Additionally, our inferred model can be naturally integrated with existing multi-armed selection strategies in the online interactive collaborating setting. Empirical studies on two real-world applications, online recommendations of movies and news, demonstrate both the effectiveness and efficiency of the proposed approach.

\end{abstract}


\maketitle

\section{Introduction} \label{sec:introduction}
The overwhelming amount of data necessitates an efficient online interactive recommendation system where the online users constantly interact with the system, and user feedback is instantly collected for improving recommendation performance.
Online interactive recommender systems are challenged to immediately suggest the consumers with appropriate items (e.g., movies, news articles) according to the current context that includes both the consumer and item content information.
The systems try to continuously maximize the consumers' satisfaction over the long run.
To achieve this goal, it becomes a critical task for recommender systems to track the consumer preferences instantly and to recommend interesting items to the users from a large item repository.

However, identifying the appropriate match between the consumer preferences and the target items is quite difficult for recommender systems due to several existing practical challenges.
One challenge is the well-known \textit{cold-start} problem since a significant number of users/items might be completely new to the system, that is, they may have no consumption history at all.
This problem makes recommender systems ineffective unless additional information about both items and users is collected~\cite{zeng2016online}, \cite{chang2015space}.
The second challenge is that most recommender systems typically assume that the entire set of contextual features with respect to both users and items can be accessed for users' preference inference. 
Due to a number of reasons -{}- privacy or sampling constraints are among them -{}- it is challenging to obtain all relevant features ahead of time, thus rendering many factors unobservable to the recommendation algorithms.

In the first challenge, an \textit{exploration} or \textit{exploitation} dilemma is identified in the aforementioned setting. A tradeoff between two competing goals needs to be considered in recommender systems: maximizing user satisfaction using their consumption history, while gathering new information for improving the goodness of match between user preference and items~\cite{li2010contextual}.
 This dilemma is typically formulated as a multi-armed bandit problem where each arm corresponds to one item.
 The recommendation algorithm determines the strategies for selecting an arm to pull according to the contextual information at each trial.
 Pulling an arm indicates that the corresponding item is recommended. When an item matches the user preference (e.g., a recommended news article or movie is consumed), a reward is obtained; otherwise, no reward is provided. The reward information is fed back to the algorithm to optimize the strategies.
 The optimal strategy is to pull the arm with the maximum expected reward with respect to the historical interaction on each trial, and then to maximize the total accumulated reward for the whole series of trials.

Collaborative filtering (CF), an early recommendation technique, is widely applied in recommender systems to address the second challenge, mentioned above~\cite{schafer2007collaborative},\cite{bennett2007netflix},\cite{koren2009matrix}.
CF has gained its popularity due to its advantage over other recommendation techniques, where CF requires no extra information about items or users for recommendation but only requires users' historical ratings on items~\cite{kawale2015efficient,herlocker1999algorithmic}. 
Further, considering both aforementioned challenges simultaneously aggravates the difficulties when recommending items.

Recently, an online interactive collaborative filtering system has been suggested~\cite{zhao2013interactive, kawale2015efficient} adopting both techniques, multi-armed bandit and collaborative filtering.
Typically, the collaborative filtering task is formulated as a matrix factorization problem. Matrix factorization derives latent features for both users and items from the historical interaction records.
It assumes that a user's preference (i.e., rating) on a given item can be predicted by considering items' and users' latent feature vectors.
Based on this assumption, multi-armed bandit policies make use of the predicted reward (i.e., user's preference) for arm (i.e., item) selection.
The feedback occurring between the given user and arm is used to update the user's and arm's latent vectors, without impacting the inference of other arms' latent vectors by supposing arms are independent from each other.

However, the assumption about the independency among arms rarely holds in real-world applications. For example, in the movie recommendation scenario, the movies correspond to the arms. The dependent arms (i.e., movies) typically share similar latent topics (e.g., science fiction movies, action movies, etc.), and are likely to receive similar rewards (i.e., ratings or feedbacks) from users. Intuitively, the dependencies among arms can be utilized for reward prediction improvement and further facilitated the maximization of users' satisfaction in the long run.

In this paper, we introduce an interactive collaborative topic regression model that utilizes multi-armed bandit algorithms with dependent arms for the item recommendations to the target user. A sequential online inference method is proposed to learn the latent parameters and infer the latent states. We adopt a generative process based on topic model to explicitly formulate the arm dependencies as the clusters on arms, where dependent arms are assumed to be generated from the same cluster.
Every time an arm is pulled, the feedback is not only used for inferring the involved user and item latent vectors, but it is also employed to update the latent parameters with respect to the arm's cluster. The latent cluster parameters further help with reward prediction for other arms in the same cluster. The fully adaptive online inference strategy of particle learning~\cite{carvalho2010particle} allows our model to effectively  capture the arm dependencies. In addition, the learnt parameters can be naturally integrated into existing multi-arm selection strategies, such as \textit{UCB} and \textit{Thompson sampling}.
We conduct empirical studies on two real-world applications, recommendations of movie and news.
The experimental results demonstrate the effectiveness of our proposed approach.

The rest of this paper is organized as follows.
In Section~\ref{sec:relatedwork}, we provide a brief summary of prior work relevant to collaborative filtering, collaborative topic model, multi-armed bandit and the online inference with particle learning.
We formulate the problem in Section~\ref{sec:formulation}.
The solution to the problem is presented in Section \ref{sec:solution}.
Extensive empirical evaluation results are reported in Section \ref{sec:experiment}. Finally, Section \ref{sec:conclusion} concludes the paper.

\section{Related Work} \label{sec:relatedwork}
In this section, we highly review the existing works that related to our approach.

\subsection{Interactive Collaborative Filtering}
Collaborative filtering (CF) methods play a key role in recommender systems that recommend items to a target user only based on users' historical ratings of items. There are two primary categories of CF technologies: the memory-based methods~\cite{sarwar2001item, herlocker1999algorithmic} 
and the model-based methods~\cite{salakhutdinov2007probabilistic, salakhutdinov2008bayesian}.

Matrix factorization (MF), one of the model-based methods gained popularity due to the Netflix Prize and other recommendation competitions. A significant variety of MF-based methods are proposed.
Probabilistic Matrix Factorization (PMF)~\cite{salakhutdinov2007probabilistic} models the ratings as products of users' and items' latent features considering Gaussian observation noise.
Bayesian Probabilistic Matrix Factorization (BPMF)~\cite{salakhutdinov2008bayesian} presents a fully Bayesian treatment of the PMF model with priors controlling model complexity automatically. 
However, one of the key challenges existing in the aforementioned methods is an effective recommendation for a new user/item in recommender systems, also referred to as a cold-start problem~\cite{li2010contextual}. Another challenge is to sequentially recommend items to the target user in an online setting and instantly adapt the up-to-date feedback to refine the recommendation predictions. Multi-armed bandit algorithms~\cite{tokic2010adaptive,auer2002using,auer2002nonstochastic,chapelle2011empirical} are widely applied to address the cold-start problem and balance the tradeoff between \textit{exploration} and \textit{exploitation} in online recomemnder systems~\cite{li2010contextual,tang2015personalized,zeng2016online}.
Interactive Collaborative Filtering (ICF)~\cite{zhao2013interactive} tackles these problems in a partially online setting leveraging PMF framework and bandit algorithms. Rao-Blackwellized particle based on Thompson sampling~\cite{chapelle2011empirical} is proposed for a fully online MF recommendation~\cite{kawale2015efficient}.

However, most of the prior bandit problems focus on independent arms. A delicate framework~\cite{pandey2007multi} is developed to study the bandit problems with dependent arms, where the dependencies are in the form of a generative model on clusters of arms. Recently, several attempts~\cite{wang2011collaborative,purushotham2012collaborative,wangbayesian} have been made by researchers to integrate CF based on MF and topic modeling~\cite{blei2003latent} in an offline setting. The central idea behind these methods is to leverage topic modeling to capture item's content information in latent topic space to soft cluster items, and focus on combining the merits of both traditional CF and probabilistic topic modeling. Different from the above mentioned work, we proposed an ICTR (\underline{I}nteractive \underline{C}ollaborative \underline{T}opic \underline{R}egression) model that adopts a generative process based on topic modeling to explicitly formulate the arm dependencies as the clusters on arms in an online setting. 

\subsection{Sequential Online Inference}
Sequential online inference algorithm is well developed to learn the latent unknown parameters and infer the latent states in our model. Sequential monte carlo sampling~\cite{halton1962sequential, doucet2001introduction} and particle learning~\cite{carvalho2010particle} are both popular sequential learning methods. Sequential Monte Carlo (SMC) methods, also known as particle filters, are a set of simulation-based methods for computing the posterior distribution to address the filtering problem~\cite{doucet2000sequential}. 
Particle learning (PL) provides state filtering, sequential parameter learning and smoothing in a large class of state space models~\cite{carvalho2010particle,zeng2016onlinebigdata}, which is an approach approximating the sequence of filtering and smoothing distributions in light of parameter uncertainty for a wide class of state space models.  The central idea behind PL is the creation of a particle algorithm that directly samples from the particle approximation to the joint posterior distribution of states and conditional sufficient statistics for fixed parameters in a fully-adapted re-sample propagate framework.

\section{Problem Formulation}
\label{sec:formulation}

In this section, we provide a mathematical formulation of the interactive collaborative filtering problem as a multi-armed bandit problem.
Then we come up with a new generative model describing explicitly the dependency among arms (i.e. items).
 
A glossary of notations mentioned in this paper is summarized in Table~\ref{tab:notations}.

\begin{table}[!h]
\vspace{-0.15in}
	\centering
	\caption{Important Notations}
    \scalebox{0.9}{
	\begin{tabular}{lp{7.5cm}}
		\toprule
		\textbf{Notation} & \textbf{Description}\\
		\midrule
		$M, N$       & number of rows (users) and columns (items).  \\
		$\mathbf{R} \in \mathcal{R}^{M \times N}$   & the rating matrix. \\
        $\mathbb{S}(t)$ & the sequence of $(n(t-1), r_{m,n(t-1)})$ observed until time $t$.\\
        $n(t)$   & the recommended item index in the $t$-th iteration. \\
        $r_{m,t}$  & the reward of the $m$-th user by pulling the given item in the $t$-th iteration. \\
		$y_{m,t}$  & the predicted rating for the $m$-th user over given item in the $t$-th iteration. \\
        $\pi$ & the policy to recommend items sequentially. \\
        $R_\pi$ & the cumulative rating (reward) of the policy $\pi$. \\
        $K$     & the number of topics and the number of dimensions for latent vectors. \\
        $\mathcal{P}_{k}$  & the set of particles for the item $k$ and $\mathcal{P}^{(i)}_{k}$.\\
        $\mathbf{p}_m \in \mathcal{R}^K$ & the latent feature vector for the $m$-th user. \\
        $\mathbf{q}_n \in \mathcal{R}^K$ & the latent feature vector for the $n$-th item. \\
        $\mathbf{\Phi}_k \in \mathcal{R}^{N}$ & the item distribution of the $k$-th topic. \\
        $z_{m,t}$ & the latent topic of the $m$-th user in the $t$-th iteration. \\
        $x_{m,t}$ & the selected item of the $m$-th user in the $t$-th iteration.\\
        $\lambda$ & Dirichlet priors over topics for topic model. \\
        $\eta$ & Dirichlet priors over items for topic model. \\
        $\sigma^2_n$ & the variance of rating prediction. \\
        $\alpha$, $\beta$ & the hyper parameters determine the distribution of $\sigma^2_n$. \\
        $\mathbf{\mu}_\mathbf{q}$, $\mathbf{\Sigma}_\mathbf{q}$ & the hyper parameters determine the gaussian distribution of $\mathbf{q}_n$. \\
		$\xi$ & the observation noise of the rating \\
        \bottomrule
	\end{tabular}
    }
	\label{tab:notations}
	\vspace{-0.15in}
\end{table}

\subsection{Basic Concepts and Terminologies}
\label{subsec:base}
Assume that there are $M$ users and $N$ items in the system. The preferences of the users for the items are recorded by a partially observable matrix $\mathbf{R}=\{r_{m,n}\}\in \mathcal{R}^{M\times N}$, where the rating score $r_{m,n}$ indicates how the user $m$ would like the item $n$. The basic collaborative filtering task is to predict the unknown rating score in light of the observed rating scores in $\mathbf{R}$. However, it is very challenging to fulfill the task in practice due to the high dimensionality and sparsity of the rating matrix. Matrix factorization addresses this challenge by mapping each user $m$ and item $n$ to the latent feature vectors $\mathbf{p}_m \in \mathcal{R}^{K}$ and $\mathbf{q}_n \in \mathcal{R}^{K}$ in a shared low $K$-dimension space (typically, $K \ll M, N$). It assumes that the rating $r_{m,n}$ can be predicted by
\begin{equation}
  \hat{r}_{m,n} = \mathbf{p}_m^\intercal\mathbf{q}_n.
  \label{eq:pq}
\end{equation}
Therefore, the latent features $\{\mathbf{p}_m\}$ and $\{\mathbf{q}_n\}$ can be learned by minimizing the prediction error for all observed ratings in $\mathbf{R}$, while each unobserved rating value can be estimated using Equation~(\ref{eq:pq}) with its corresponding latent features learned by matrix factorization.
In practice, since the feedback (i.e., rating scores) from users is received over time, the recommender system is required to address the collaborative filtering problem in an interactive mode.
The recommender system is referred to as an interactive recommender system.

In an interactive recommender system, a user $m$ constantly arrives to interact with system over time. At each time $t \in [1,T]$, the system, according to the observed rating history, recommends an item $n(t)$ to the corresponding user $m$. After consuming the item $n(t)$, the feedback (i.e., rating) $r_{m,n(t)}$ from the user $m$ is collected by the system and further utilized to update the recommendation model for the next item delivery. The interactive recommendation process involves a series of decisions over a finite but possibly unknown time horizon $T$. Accordingly, we formulate the interactive recommendation process as a multi-armed bandit problem, where each item corresponds to an arm. Pulling an arm indicates that its corresponding item is being recommended, and the rating score is considered as the reward received after pulling the corresponding arm.

Let $\mathbb{S}(t)$ be the available information at time $t$ collected by the system for the target user $m$,
\begin{equation}
 \begin{split}
 \mathbb{S}(t) = \{(n(1), r_{m,n(1)}),\dots,(n(t-1), r_{m, n(t-1)})\}.
 \end{split}
 \end{equation}
A policy $\pi$ is defined as a function and used to select an arm based on the current cumulative information $\mathbb{S}(t)$,
\begin{equation}
 \begin{split}
 n(t) = \pi(\mathbb{S}(t)).
 \end{split}
 \end{equation}
The total reward received by the policy $\pi$ after $T$ iterations is
\begin{equation}
  R_\pi = \sum_{t=1}^{T}{r_{m,\pi(\mathbb{S}(t))}}.
  \label{eq:totalReward}
\end{equation}
The optimal policy $\pi^*$ is defined as the one with maximum accumulated expected reward after $T$ iterations,
\begin{equation}
 \begin{split}
 \pi^* = \arg\max_{\pi}{\mathbb{E}(R_\pi)} = \arg\max_{\pi}\sum_{t=1}^{T}\mathbb{E}(r_{m,\pi(\mathbb{S}(t))}|t).
 \end{split}
 \label{eq:optPolicy}
 \end{equation}
 Therefore, our goal is to identify a good policy for maximizing the total reward.
 Herein we use reward instead of regret to express the objective function, since  maximization of the cumulative rewards is equivalent to minimization of regret during the $T$ iterations~\cite{zhao2013interactive}.

Before selecting one arm at time $t$, a policy $\pi$ typically learns a model to predict the reward for every arm according to the historical accumulated information $\mathbb{S}(t)$.
The reward prediction helps the policy $\pi$ make decisions to increase the total reward.


In the latent factor model~\cite{salakhutdinov2007probabilistic,salakhutdinov2008bayesian}, the rating is estimated by a product of user and item feature vectors $\mathbf{p}_m$ and $\mathbf{q}_n$ in Equation ~(\ref{eq:pq}). 
From the probabilistic perspective, probabilistic matrix factorization introduces an observation noise $\xi$, a zero-mean Gaussian noise with variance $\sigma^2$~(i.e., $\xi \sim \mathcal{N}(0, \sigma^2)$), to the rating prediction function given in Equation~(\ref{eq:pq}). The derived rating prediction is as follows:
\begin{equation}
 \begin{split}
 r_{m,n} = \mathbf{p}_m^{\intercal}\mathbf{q}_n + \xi.
 \end{split}
 \label{eq:rewardPrediction}
 \end{equation}
In this setting, our objective function in Equation~(\ref{eq:optPolicy}) is re-formulated as:
\begin{equation}
  \begin{split}
 \pi^* = \arg\max_{\pi}\sum_{t=1}^{T}\mathbb{E}_{\mathbf{p}_m,\mathbf{q}_{\pi(\mathbb{S}(t))}}(\mathbf{p}_m^\intercal\mathbf{q}_{\pi(\mathbb{S}(t))}|t).
 \end{split}
 \label{eq:pmfOptPolicy}
\end{equation}
Consequently, the goal of the interactive recommender system is reduced to optimization of the objective function in Equation~(\ref{eq:pmfOptPolicy}).

\textbf{Thompson Sampling,} one of earliest heuristics for the bandit problem~\cite{chapelle2011empirical}, belongs to the probability matching family.
Its main idea is to randomly allocate the pulling chance according to the probability that an arm gives the largest expected reward at a particular time $t$.
Based on the objective function in Equation~(\ref{eq:pmfOptPolicy}), the probability of pulling arm $n$ can be expressed as follows:
\begin{equation}
  \begin{split}
    p(n(t) = n) = \int & {\mathbb{I}[\mathbb{E}(r_{m,n}|\mathbf{p}_m,\mathbf{q}_n)]}) = \max_{i}{\mathbb{E}(r_{m,i}|\mathbf{p}_m,\mathbf{q}_i)}] \\ & p(\mathbf{p}_m,\mathbf{q}_n|t)d\mathbf{p}_md\mathbf{q}_n.
  \end{split}
\end{equation}
At each time $t$, the Thompson sampling algorithm samples both the user and item feature vectors together from their corresponding distributions, and then selects the item that leads to the largest reward expectation.
Therefore, using the Thompson sampling strategy, the item selection function is defined as:
\begin{equation}
  n(t) = \arg\max_{n}{(\mathbf{\tilde{p}}_m^\intercal\mathbf{\tilde{q}}_n|t)},
  \label{eq:funSelection}
\end{equation}
where $\mathbf{\tilde{p}}_m$ and $\mathbf{\tilde{q}}_n$ denote the sampled feature vectors for user $m$ and item $n$, respectively.

To accomplish the Thompson sampling, it is critical to model the random variable $\mathbf{p}_m$ and $\mathbf{q}_n$ using distributions, where the latent feature vectors can be easily sampled and the feedback at every time can be reasonably integrated.
Most of the previous studies suppose a Gaussian prior for both user and item feature vectors, with an assumption that the items are independent from each other~\cite{zhao2013interactive,kawale2015efficient}.
However, this assumption rarely holds in real applications.
In the following section, we explicitly formulate the dependent arms with a generative model.
\subsection{Modeling the Arm Dependency} \label{subsec:dependence-arm}

In light of the fact that similar items (i.e., arms) are likely to receive similar feedback (i.e., rewards)
, we assume that a dependency existing among similar items.
The dependencies among items can be further leveraged to improve the users' preferences inference on a particular item even if the item has little historical interaction data in the recommender system.
The challenge here lies in how to sequentially infer the arms' dependencies as well as users' preferences simultaneously, providing the feedback over time.

In our work, the dependencies among arms are expressed in the form of the clusters of arms, where the dependent arms fall into the same cluster.
In order to explore the dependencies in the multi-armed bandit setting, Latent Dirichlet Allocation (abbr., LDA~\cite{blei2003latent}), a generative statistic model for topic modeling, is adopted to construct the arms' clusters.
We propose the ICTR (\underline{I}nteractive \underline{C}ollaborative \underline{T}opic \underline{R}egression) model to infer the clusters of arms as well as arm selection. 



\begin{figure}[!htb]
    \centering
    \scalebox{0.6}{
        \includegraphics {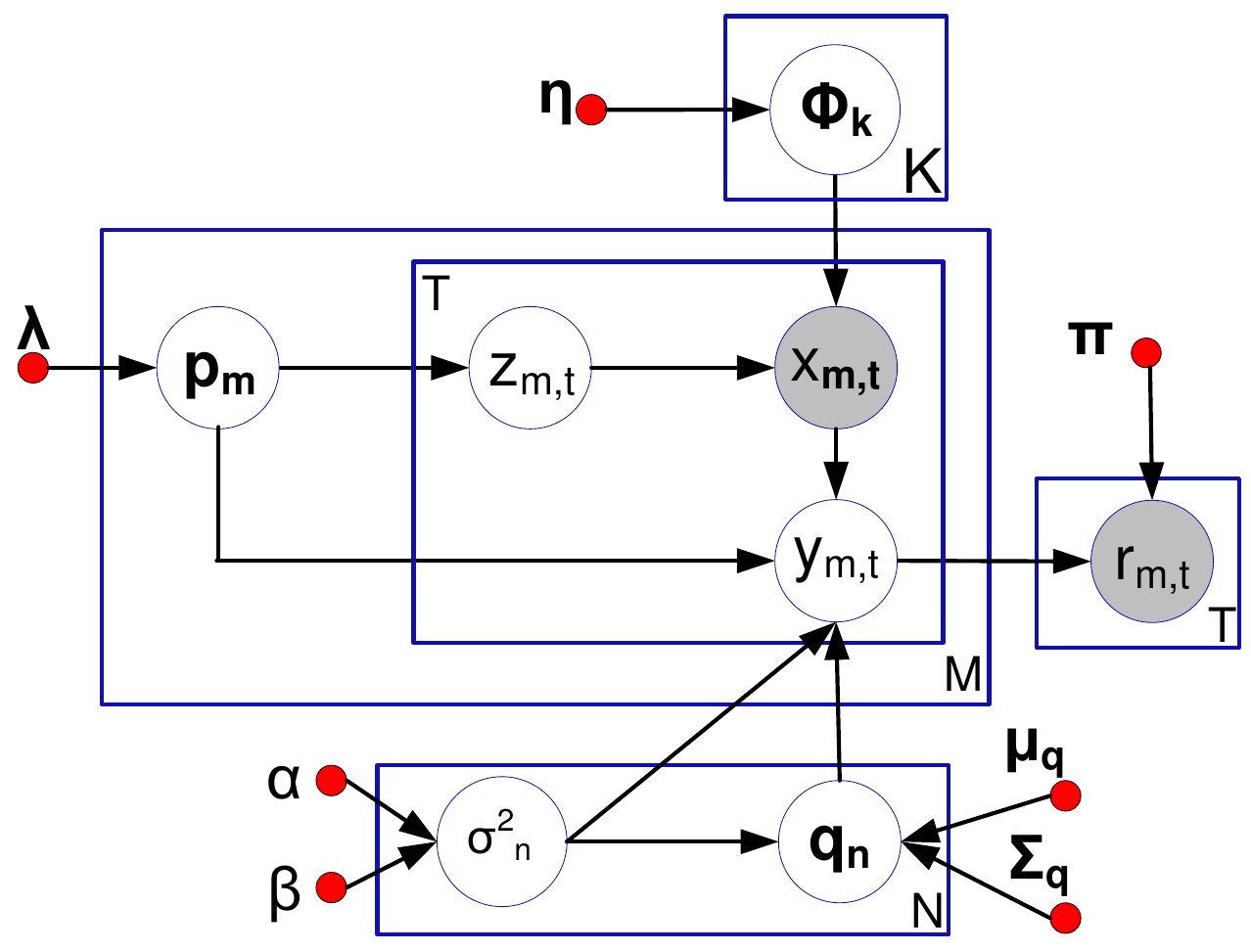}
    }
    \caption{The graphic model for the ICTR model. Random variable is denoted as a circle. The circle with filled color denotes the observed random variable. Red dot represents a hyper parameter.}
    \label{fig:model}
\end{figure}

The main idea of our model is to treat an item $n$ as a word, and consider a user $m$ as a document.
All the items rated by a user indicate the hidden preferences of the user, analogous to the scenario in topic modeling where the words contained in a document imply its latent topics.
Specifically, let $K$ be the number of latent aspects (i.e., topics or clusters) the users' concern when consuming items.
We assume that $\mathbf{p}_m\in\mathcal{R}^K$ corresponds to the latent vector for user $m$, where the $k$-th component of $\mathbf{p}_m$ suggests the user's preference over the $k$-th aspect of items.
Further, $\mathbf{q}_n\in\mathcal{R}^K$ is supposed to be the latent vector for the item $n$, and the $k$-th component value of $\mathbf{q}_n$ represents the it belongs to the $k$-th cluster.
The rating score $r_{m,n}$, given by user $m$ after consuming item $n$, is assumed to be the inner product of $\mathbf{p}_m$ and $\mathbf{q}_n$. 
By linking to the topic model, a generative process for user ratings is accordingly introduced and presented in Figure~\ref{fig:model}.

Based on the above description, the user latent vector $\mathbf{p}_m$ is assumed to follow a Dirichlet prior distribution with a predefined hyper parameter $\mathbf{\lambda}$, shown in Equation~\ref{eq:pmPrior}. 
\begin{equation}
 \mathbf{p}_m|\mathbf{\lambda} \sim Dir(\mathbf{\lambda}).
 \label{eq:pmPrior}
 \end{equation}
As presented in Equation~\ref{eq:rewardPrediction}, we denote $\mathbf{\sigma}^2$ as the variance of the noise for reward prediction, and we assume $\sigma^2_n$ is drawn from the Inverse Gamma (abbr., $\mathcal{IG}$) distribution shown in Equation~\ref{eq:deltaPrior}.
\begin{equation}
 p(\sigma^2_n|\alpha, \beta) = \mathcal{IG}(\alpha, \beta),
 \label{eq:deltaPrior}
\end{equation}
where $\alpha$ and $\beta$ are the predefined hyper parameters for the Inverse Gamma distribution.

Given $\mathbf{\sigma}^2_n$, the item latent vector $\mathbf{q}_n$ is generated by a Gaussian prior distribution as follows:
\begin{equation}
\mathbf{q}_{n}|\mathbf{\mu}_\mathbf{q}, \mathbf{\Sigma}_\mathbf{q}, \sigma^2_n \sim \mathcal{N}(\mathbf{\mu}_\mathbf{q}, \sigma^2_n\mathbf{\Sigma}_\mathbf{q}),
 \label{eq:itemDistribution}
 \end{equation}
where $\mathbf{\mu}_\mathbf{q}$ and $\mathbf{\Sigma}_\mathbf{q}$ are predefined hyper parameters.

Further, let $\mathbf{\Phi}_k\in\mathcal{R}^N$ be the item distribution for the topic $k$.
Similar to $\mathbf{p}_m$, Dirichlet distribution is specified as the prior of $\mathbf{\Phi}_k$ shown in Equation~\ref{eq:phiPrior}
\begin{equation}
  \mathbf{\Phi}_k|\eta \sim Dir(\eta),
  \label{eq:phiPrior}
\end{equation}
where $\eta\in\mathcal{R}^N$ is the hyper parameter.

When user $m$ arrives to interact with the recommender system at time $t$, one of $K$ topics, denoted as $z_{m,t}$, is first selected according to the user's latent preference $\mathbf{p}_m$, indicating that the user $m$ shows interest in the topic $z_{m,t}$ at this moment.
Accordingly, $z_{m,t}$ is supposed to follow a multinomial distribution governed by $\mathbf{p}_m$ as follows,
\begin{equation}
  z_{m,t}|\mathbf{p}_m \sim Mult(\mathbf{p}_m).
  \label{eq:zDistr}
\end{equation}

Without loss of generality, we assume $z_{m,t} = k$, then the item distribution for the topic $k$ (i.e., $\mathbf{\Phi}_k$) is used for generating the item $x_{m,t}$ recommended to the user $m$ at time $t$.
We assume the random variable $x_{m,t}$ follows the multinominal distribution ruled with $\mathbf{\Phi}_k$, i.e.,
\begin{equation}
  x_{m,t}|\mathbf{\Phi}_k \sim Mult(\mathbf{\Phi}_k).
  \label{eq:xDistr}
\end{equation}

Without loss of generality, item $n$ is assumed to be selected by user $m$ at time $t$ (i.e., $x_{m,t}=n$) where the latent vector corresponding to item $n$ is $\mathbf{q}_n$.  Let $y_{m,t}$ be the predicted reward (i.e., rating), given by user $m$ at time $t$. The predicted reward $y_{m,t}$ can be inferred by
\begin{equation}
  y_{m,t} \sim \mathcal{N}(\mathbf{p}_m^\intercal\mathbf{q}_n, \sigma^2_n).
  \label{eq:predictedReward}
\end{equation}
By Equation~\ref{eq:predictedReward}, the rewards of different items are predicted. Based on the predicted rewards, the policy $\pi$ selects an item and recommends it to the user $m$, considering the trade-off between exploitation and exploration. After consuming the recommended item, the recommender system receives the actual reward $r_{m,t}$ from the user. The objective of the model is to maximize the expected accumulative rewards in the long run as described in Equation~\ref{eq:optPolicy}.

In this section, taking the clusters of arms into account, we formally introduce our ICTR model, which integrates matrix factorization with topic modeling under the multi-armed bandit setting. We develop our solution to infer the ICTR model from a Bayesian perspective and the solution is presented in the following section.

\section{Methodology and Solution}\label{sec:solution}
In this section, we present the methodology for online inferences of the interactive collaborative topic regression model.

The posterior distribution inference involves five random variables, i.e., $\mathbf{p}_{m}$,
$z_{m,t}$, $\mathbf{\Phi}_{k}$, $\mathbf{q}_n$, and ${\sigma^2_n}$. According to the graphical model in Figure~\ref{fig:model}, the five random variables belong to two categories: parameter random variable and latent state random variable.
$\mathbf{\Phi}_{k}$, $\mathbf{p}_m$, $\mathbf{q}_n$, and ${\sigma^2_n}$ are parameter random variables since they are assumed to be fixed but unknown, and their values do not change with time.
Instead, $z_{m,t}$ is referred to as a latent state random variable since it is not observable and its value is time dependent.
After pulling the arm $n(t)$, where $n(t) = x_{m,t}$ according to Equation~\ref{eq:xDistr} at time $t$, a reward is observed as $r_{m,t}$.
Thus, $x_{m,t}$ and $r_{m,t}$ are referred to as observed random variables.

Our goal is to infer both latent parameter variables and latent state random variable to sequentially fit the observed data at time $t-1$, and predict the rewards for arm selection with respect to the coming user at time $t$. However, since the inference of our model cannot be conducted by a simple closed-form solution, we adopt the sequential sampling-based inference strategy that is widely used in sequential Monte Carlo sampling~\cite{smith2013sequential}, particle filtering~\cite{djuric2003particle}, and particle learning~\cite{carvalho2010particle} to learn the distribution of both the parameter and the state random variables.
Specifically, a particle learning method that allows both state filtering and sequential parameter learning simultaneously is a perfect solution to our proposed model inference.

In order to develop the solution based on the particle learning, we first define the particle as follows.
\begin{definition}[Particle] A particle for predicting the reward $y_{m,t}$ is a container that maintains the current status information for both user $m$ and item $x_{m,t}$. The status information comprises of random variables such as $\mathbf{p}_{m}$,
${\sigma^2_n}$, $\mathbf{\Phi}_{k}$, $\mathbf{q}_n$, and $z_{m,t}$, as well as the hyper parameters of their corresponding distributions, such as $\mathbf{\lambda}$, $\alpha$, $\beta$, $\mathbf{\eta}$, $\mathbf{\mu}_\mathbf{q}$ and $\mathbf{\Sigma_q}$.
\end{definition}
In particle learning, each particle corresponds to a sample for modeling inference status information.
At each time stamp, the particles are re-sampled according to their fitness to the current observable data.
Then, the re-sampled particles are propagated to new particles and obtain the status information for the next time stamp.

In the following subsections, we develop our solution based on particle learning.

\subsection{Re-sample Particles with Weights}
At time $t-1$, a fixed-size set of particles is maintained for the reward prediction for each arm $n(t-1)$ given the user $m$. We denote the particle set at time $t-1$ as $\mathcal{P}_{m,n(t-1)}$ and assume the number of particles in $\mathcal{P}_{m,n(t-1)}$ is $B$.
Let $\mathcal{P}_{m,n(t-1)}^{(i)}$ be the $i^{th}$ particles given both user $m$  and the item $n(t-1)$ at time $t-1$, where $1\le i\le B$. Each particle $\mathcal{P}_{m,n(t-1)}^{(i)}$ has a weight, denoted as $\rho^{(i)}$,  indicating its fitness for the new observed data at time $t$. Note that $\sum_{i=1}^{B}{\rho^{(i)}} = 1$. The fitness of each particle $\mathcal{P}_{m,n(t-1)}^{(i)}$ is defined as the likelihood of the observed data $x_{m,t}$ and $r_{m,t}$. Therefore,
\begin{equation}
  \rho^{(i)} \varpropto p(x_{m,t}, r_{m,t}|\mathcal{P}_{m,n(t-1)}^{(i)}).
\label{eq:computeWeights}
\end{equation}
Further, $y_{m,t}$ is the predicted value of $r_{m,t}$. The distribution of $y_{m,t}$, determined by $\mathbf{p}_m$, $\mathbf{q}_n$, $z_{m,t}$, $\mathbf{\Phi}_k$ , and  $\sigma^2_n$, described in Section~\ref{subsec:dependence-arm}.

Therefore, we can compute $\rho^{(i)}$ as proportional to the density value given $y_{m,t} = r_{m,t}$ and $x_{m,t} = n$.
Thus, we obtain
\begin{equation*}
    \begin{split}
   \rho^{(i)} \varpropto & \sum_{z_{m,t}=1}^{K}\{\mathcal{N}(r_{m,t}|(\mathbf{p}_{m}^\intercal\mathbf{q}_{n},\sigma^2_{n}) \\  &\bullet p(z_{m,t}=k, x_{m,t} = n|\mathcal{P}_{m,n(t-1)}^{(i)})\},
 \end{split}
 \end{equation*}
 where
 \begin{equation}
 \begin{split}
   &p(z_{m,t}=k, x_{m,t} = n|\mathcal{P}_{m,n(t-1)}^{(i)}) \\
   &=\iint_{\mathbf{p}_m,\mathbf{\Phi}_k}p(z_{m,t}=k, x_{m,t} = n,\mathbf{p}_m,\mathbf{\Phi}_k|\lambda,\eta)d\mathbf{p}_m,d\mathbf{\Phi}_k \\
   &=\int_{\mathbf{p}_m}Mult(z_{m,t}=k|\mathbf{p}_m)Dir(\mathbf{p}_m|\lambda)d\mathbf{p}_m \\ &\bullet\int_{\mathbf{\Phi}_k}Mult(x_{m,t}=n|\mathbf{\Phi}_k)Dir(\mathbf{\Phi}_k|\eta)d\mathbf{\Phi}_k \\
   &=E(\mathbf{p}_{m,k}|\lambda)\bullet E(\mathbf{\Phi}_{k,n}|\eta).
 \end{split}
 \label{eq:zInfer}
 \end{equation}
 Thus, we have:
 \begin{equation}
   \rho^{(i)} \varpropto \sum_{z_{m,t}=1}^{K}\{\mathcal{N}(\mathbf{r}_{m,t}|(\mathbf{p}_{m}^\intercal\mathbf{q}_{n},\sigma^2_{n})\bullet E(\mathbf{p}_{m,k}|\lambda)\bullet E(\mathbf{\Phi}_{k,n}|\eta)\},
   \label{eq:weight}
 \end{equation}
 where $E(\mathbf{p}_{m,k}|\lambda)$ and $E(\mathbf{\Phi}_{k,n}|\eta)$ represent the conditional expectations of $\mathbf{p}_{m,k}$ and $\mathbf{\Phi}_{k,n}$ given the observed reward $\lambda$ and $\eta$ of $\mathcal{P}_{m,n(t-1)}^{(i)}$ . The expectations can be inferred by $E(\mathbf{p}_{m,k}|\lambda) = \frac{\lambda_{k}}{\sum_{k=1}^{K}{\lambda_{k}}}$ and $ E(\mathbf{\Phi}_{k,n}|\eta) = \frac{\eta_{k,n}}{\sum_{n=1}^{N}{\eta_{k,n}}}.$

Before updating any parameters, a re-sampling process is conducted. We replace the particle set $\mathcal{P}_{m,n(t-1)}$ with a new set $\mathcal{P}_{m,n(t)}$, where $\mathcal{P}_{m,n(t)}$ is generated from $\mathcal{P}_{m,n(t-1)}$ using sampling with replacement based on the weights of particles. Then sequential parameter updating is based on $\mathcal{P}_{m,n(t)}.$

 \subsection{Latent State Inference}
 Provided with the new observation $x_{m,t}$ and $r_{m,t}$ at time $t$, the random state $z_{m,t}$ can be one of $K$ topics, and the posterior distribution of $z_{m,t}$ is shown as follows:
 \begin{equation}
   z_{m,t}|x_{m,t},r_{m,t},\mathcal{P}_{m,n(t-1)}^{(i)} \sim Mult(\mathbf{\theta}),
   \label{eq:sampleZ}
 \end{equation}
 where $\mathbf{\theta}\in\mathcal{R}^K$ is the parameter specifying the multinominal distribution. According to Equation~\ref{eq:zInfer}, since
 \begin{equation*}
 p(z_{m,t}|x_{m,t}, r_{m,t},\lambda,\eta) \varpropto p(z_{m,t},x_{m,t}|r_{m,t},\lambda,\eta),
 \end{equation*}
$\theta$ can be computed by $\theta_k \varpropto E(\mathbf{p}_{m,k}|r_{m,t},\lambda)\bullet E(\mathbf{\Phi}_{k,n}|r_{m,t},\eta).$
 Further, $E(\mathbf{p}_{m,k}|r_{m,t},\lambda)$ and $E(\mathbf{\Phi}_{k,n}|r_{m,t},\eta)$ can be obtained as follows,
 \begin{equation}
   \begin{split}
     E(\mathbf{p}_{m,k}|r_{m,t},\lambda) = \frac{\mathcal{I}(z_{m,t}=k)r_{m,t} + \lambda_{k}}{\sum_{k=1}^{K}[\mathcal{I}(z_{m,t}=k)r_{m,t} + \lambda_{k}]},
   \end{split}
   \label{eq:expectP}
   \end{equation}
   and
   \begin{equation}
   \begin{split}
     E(\mathbf{\Phi}_{k,n}|r_{m,t},\eta) = \frac{\mathcal{I}(x_{m,t}=n)r_{m,t} + \eta_{k,n}}{\sum_{n=1}^{N}[\mathcal{I}(x_{m,t}=n)r_{m,t} + \eta_{k,n}]},
   \end{split}
   \label{eq:expectPhi}
 \end{equation}
 where $\mathcal{I}(\bullet)$, an indicator function, returns $1$ when the input boolean expression is true, otherwise returns $0$. Specifically, if $r_{m,t}\in\{0,1\}$, the value of $r_{m,t}$ indicates whether $x_{m,t}$ should be included in the preferred item list of the user $m$. If $r_{m,t}\in [0,+\infty)$, the value of $r_{m,t}$ implies how the user $m$ likes the item $x_{m,t}$. Therefore, our solution can effectively handle the non-negative rating score at different scales.
\subsection{Parameter Statistics Inference}
At time $t-1$, the sufficient statistics for the parameter random variables ($\mathbf{q}_n$, ${\sigma^2_n}$, $\mathbf{p}_{m}$,
$\mathbf{\Phi}_{k}$) are ($\mathbf{\mu}_\mathbf{q}$, $\mathbf{\Sigma}_\mathbf{q}$, $\alpha$, $\beta$, $\mathbf{\lambda}$, $\mathbf{\eta}$). Assume $\mathbf{\mu}_\mathbf{q}'$, $\mathbf{\Sigma}_\mathbf{q}'$, $\alpha'$, $\beta'$, $\mathbf{\lambda}'$, $\mathbf{\eta}'$ are the sufficient statistics at time $t$, which are updated on the basis of both the sufficient statistics at time $t-1$ and the new observation data (i.e, $x_{m,t}$ and $r_{m,t}$). The sufficient statistics for parameters are updated as follows:
 \begin{equation}
 \begin{split}
  & \mathbf{\Sigma}_{\mathbf{q}_n}' = (\mathbf{\Sigma}_{\mathbf{q}_n}^{-1} + \mathbf{p}_m\mathbf{p}_m^{\intercal})^{-1} \\
  & \mathbf{\mu}_{\mathbf{q}_n}' = \mathbf{\Sigma}_{\mathbf{q}_n}'(\mathbf{\Sigma}_{\mathbf{q}_n}^{-1}\mathbf{\mu}_{\mathbf{q}_n} + \mathbf{p}_m r_{m,t}) \\
  & \alpha' = \alpha + \frac{1}{2} \\
  & \beta' = \beta + \frac{1}{2}(\mathbf{\mu}_{\mathbf{q}_n}^{\intercal} \mathbf{\Sigma}_{\mathbf{q}_n}^{-1} \mathbf{\mu}_{\mathbf{q}_n} +
  r_{m,t}^{\intercal}r_{m,t} - \mathbf{\mu}_{\mathbf{q}_n}'^{\intercal} \mathbf{\Sigma}_{\mathbf{q}_n}'^{-1} \mathbf{\mu}_{\mathbf{q}_n}') \\
  & \lambda'_{k} = \mathcal{I}(z_{m,t}=k)r_{m,t} + \lambda_{k} \\
  & \eta'_{k,n} = \mathcal{I}(x_{m,t}=n)r_{m,t} + \eta_{k,n}
 \end{split}
 \label{eq:updateParameters}
 \end{equation}

At time t, the sampling process for the parameter random variables ${\sigma^2_n}$, $\mathbf{q}_n$, $\mathbf{p}_{m}$ and
$\mathbf{\Phi}_{k}$ is summarized as below:
\begin{equation}
 \begin{split}
 & \sigma^2_n \sim \mathcal{IG}(\alpha', \beta'), \\
 & \mathbf{q}_n|\sigma^2_n \sim \mathcal{N}(\mathbf{\mu}_{\mathbf{q}_n}', \sigma^2_n\mathbf{\Sigma}_{\mathbf{q}_n}'), \\
 & \mathbf{p}_{m} \sim Dir(\lambda'), \\
 & \mathbf{\Phi}_{k} \sim Dir(\eta').
 \label{eq:sampleParameters}
 \end{split}
\end{equation}

  \subsection{Integration with Policies}\label{sec:policyIntegration}


In our interactive collaborative topic regression model, when user $m$ arrives at time $t$, the reward $r_{m,t}$ is unknown since it is not observed until one of arms $x_{m,t}$ is pulled. Without observed $x_{m,t}$ and $r_{m,t}$, the particle re-sampling, latent state inference, and parameter statistics inference for time $t$ cannot be conducted. Therefore, we utilize the latent vectors $\mathbf{p}_m$ and $\mathbf{q}_n$, sampled from their corresponding posterior distributions by Equation~\ref{eq:sampleParameters} at time $t-1$, to predict the reward for every arm. In this section, two policies for interactive collaborative filtering, based on Thompson sampling and UCB (\underline{U}pper \underline{C}onfidence \underline{B}ound), are integrated with our model.

In our model, given the user $m$, every item has $B$ independent particles. Each particle $i$ contains its latent variables and parameters, and produces an independent reward prediction $r_{m,t}^{(i)}.$

Specifically, according to Thompson sampling discussed in Section~\ref{subsec:base}, we predict the reward of pulling arm $n$ with the average value of rewards from $B$ particles. 
The policy based on Thompson sampling selects an arm $n(t)$ based on the following equation,
\begin{equation}
  n(t) =\arg\max_{n}{(\bar{r}_{m,n})},
  \label{eq:selectArmTS}
\end{equation}
where $\bar{r}_{m,n}$ denotes the average reward, i.e., $$\bar{r}_{m,n} = \frac{1}{B}\sum_{i=1}^{B}{\mathbf{p}_m^{(i)\intercal}\mathbf{q}_n^{(i)}}.$$

Moreover, the UCB policy selects an arms based on the upper bound of the predicted reward. Assuming that $$r_{m,t}^{(i)}
\sim \mathcal{N}(\mathbf{p}_m^{(i)\intercal}\mathbf{q}_n^{(i)},{\sigma^{(i)2}}),$$ the UCB based policy is developed by the mean and variance of predicted reward, i.e.,
\begin{equation}
  n(t) = \arg\max_{n}{(\bar{r}_{m,n} + \gamma\sqrt{\nu})},
\label{eq:selectArmUCB}
\end{equation}
where $\gamma \ge 0$ is a predefined threshold, and the variance is expressed as,
$$\nu = \frac{1}{B}\sum_{i}^{B}{\sigma^{(i)2}}.$$
\subsection{Algorithm}
Putting all the aforementioned inference together, an algorithm based on the interactive collaborative topic regression model is provided below.

Online inference for the interactive collaborative filtering problem starts with $\text{MAIN}$ procedure, as presented in Algorithm~\ref{alg:ictr}. As user $m$ arrives at time $t$, the $\text{EVAL}$ procedure computes a score for each arm, where we define the score as the average reward. The arm with the highest score is selected to pull. After receiving a reward by pulling an arm, the new feedback is used to update the ICTR model by the $\text{UPDATE}$ procedure. Especially in the $\text{UPDATE}$ procedure, we use the $resample$-$propagate$ strategy in particle learning~\cite{carvalho2010particle} rather than the $propagate$-$resample$ strategy in particle filtering~\cite{djuric2003particle}. With the $resample$-$propagate$ strategy, the particles are re-sampled by taking $\rho^{(i)}$ as the $i^{th}$ particle's weight, where the $\rho^{(i)}$ indicates the fitness for the observation at time $t$ given the particle at time $t-1$. The $resample$-$propagate$ strategy is considered as an optimal and fully adapted strategy, avoiding an importance sampling step.

  \begin{algorithm}[!htp]
    \caption{The algorithm for interactive collaborative topic regression model (\texttt{ICTR})}
     \label{alg:ictr}
    \begin{algorithmic}[1]
        \Procedure{main}{$B$}\Comment{main entry}
            \State {Initialize $B$ particles, i.e., $\mathcal{P}_{m,n(0)}^{(1)}...\mathcal{P}_{m,n(0)}^{(B)}$.}
            \For {$t\gets 1, T$}
            \State{User $m$ arrives for item recommendation.}
            \State {$n(t) = \arg\max_{n=1,N}{\text{EVAL}(m,n)}$}\Comment{by Eq.~\ref{eq:selectArmTS} or Eq.~\ref{eq:selectArmUCB}.}
            \State {Receive $r_{m,t}$ by rating item $n(t)$.}
            \State {\text{UPDATE}($m$, $n(t)$, $r_{m,t}$). }
            \EndFor
        \EndProcedure
        \Statex{}
        \Procedure {eval}{$m$, $n$}\Comment{get a rating score for item $n$, given user $m$.}
            \For {$i\gets 1, B$}\Comment{Iterate on each particle.}
            \State{Get the user latent vector $\mathbf{p}_m^{(i)}$.}
            \State{Get the item latent vector $\mathbf{q}_n^{(i)}$.}
            \State{Predict $i^{th}$ reward $r_{m,t}^{(i)}$.}
            \EndFor
            \State{Compute the average reward as the final reward $r_{m,t}$.}
            \State \textbf{return} {the score.}
        \EndProcedure
        \Statex{}
        \Procedure {update}{$m$, $n(t)$, $r_{m,t}$}\Comment{update the inference.}
            \For {$i\gets 1, B$}\Comment{Compute weights for each particle.}
                \State {Compute weight $\rho^{(i)}$ of particle $\mathcal{P}_{m,n(t)}^{(i)}$ by Eq.~\ref{eq:computeWeights}.}
            \EndFor
            \State {Re-sample $\mathcal{P'}_{m,n(t)}$ from $\mathcal{P}_{m,n(t)}$ according to the weights $\rho^{(i)}$s.}
            \For {$i\gets 1, B$}\Comment{Update statistics for each particle.}
                \State {Update the sufficient statistics for $z_{m,t}$ by Eq.~\ref{eq:expectP} and~\ref{eq:expectPhi}.}
                \State {Sample $z_{m,t}$ according to Eq.~\ref{eq:sampleZ}.}
                \State {Update the statistics for $\mathbf{q}_n$, ${\sigma^2_n}$, $\mathbf{p}_{m}$,
$\mathbf{\Phi}_{k}$ by Eq.~\ref{eq:updateParameters}.}
                \State {Sample $\mathbf{q}_n$, ${\sigma^2_n}$, $\mathbf{p}_{m}$,
$\mathbf{\Phi}_{k}$ according to Eq.~\ref{eq:sampleParameters}.}
            \EndFor
        \EndProcedure
    \end{algorithmic}
  \end{algorithm}

In addition, existing algorithms~\cite{zhao2013interactive,kawale2015efficient} consider all the arms independently, while our model takes the clusters of arms into account by learning the topic-related random variables (e.g., $\Phi_k$). Those topic related random variables are shared among all the arms.
\section{Empirical Study}\label{sec:experiment}

  To demonstrate the efficiency of our proposed algorithm, we conduct our experimental study over two popular real-world dataset: Yahoo! Today News and MovieLens (10M).
First, we outline the general implementation of the baseline algorithms for comparison. Second,  we start with a brief description of the data sets and evaluation method.
Finally, we show and discuss the comparative experimental results of both the proposed algorithms and the baseline algorithms, and a case study on movie topic distribution analysis of MovieLens (10M).

\subsection{Baseline Algorithms}\label{sec:baseline}
In the experiment, we demonstrate the performance of our methods by comparing them with the following baseline algorithms:

\begin{enumerate}
  \item \texttt{Random}: it randomly selects an item recommending to the target user.

  \item \texttt{$\epsilon$-greedy}($\epsilon$): it randomly selects an item with probability $\epsilon$ and selects the item of the largest predicted reward with probability $1-\epsilon$, where $\epsilon$ is a predefined parameter.

  \item \texttt{UCB}($\lambda$): it picks the item $j(t)$ with the highest rewards at time $t$ as follows:
  \begin{equation*}
     j(t) = \arg\max_{j=1,\dots,N}(\hat{\mu_i} + \lambda\sqrt{\frac{2ln(t)}{n_{i(t)}}})
  \end{equation*}
  where $n_{i(t)}$ is the number of times that item $n_i$ has been recommended until time $t$.
  \item \texttt{TS}($S_i(t), F_i(t)$): Thompson sampling described in Section~\ref{subsec:base}, randomly draws the expected reward from the Beta posterior distribution, and selects the item of the largest predicted reward. $S_i(t)$/$F_i(t)$ indicates the number of positive/negative feedback on item $i$ until time $t$.
  \item \texttt{PTS}($d$, $p$): Particle Thompson sampling for matrix factorization, approximates the posterior of the latent feature vectors by updating a set of particles. Here $d$ is the dimension of latent feature vector and $p$ is the number of particles.
\end{enumerate}

 Our methods proposed in this paper include:
 \begin{enumerate}
 	\item \texttt{ICTRTS}($d$, $p$): it denotes our proposed interactive collaborative topic regression model with \texttt{TS} bandit algorithm. Here $d$ is the dimension of latent feature vector and $p$ is the number of particles.
 	\item \texttt{ICTRUCB}($d$, $p$, $\gamma$): it indicates our proposed model with \texttt{UCB} bandit algorithm. Similar to \texttt{UCB}, $\gamma$ is given. Here $d$ is the dimension of latent feature vector and $p$ represents the number of particles.
 \end{enumerate}
\subsection{Datasets Description}
 We use two real-world datasets shown in Table~\ref{tab:datasets} to evaluate our proposed algorithms.
\begin{table}[!htb]
        \centering
        \vspace{-0.1in}
        \caption{Description of Datasets.}
       
        \begin{tabular}{|c|c|c|}
        \hline
        \textbf{Dataset}  & {\textbf{Yahoo News}}  & {\textbf{MovieLens (10M)}}\\
        \hline
        \texttt{\#users}   & 226,710    &  71,567       \\
        \hline
        \texttt{\#items}   & 652        &  10,681  \\
       \hline
        \texttt{\#ratings} & 28,041,0150  &  10,000,054  \\
       \hline
        \end{tabular}
        
        \label{tab:datasets}
\end{table}

\textbf{Yahoo! Today News:} The core task of personalized news recommendation is to display appropriate news articles on the web page for the users according to the potential interests of individuals. The recommender system often takes the instant feedback from users into account to improve the prediction of the potential interests of individuals, where the user feedback is about whether the users click the recommended article or not.
 Here, we formulate the personalized news recommendation problem as an instance of bandit problem, where each item corresponds to a news article and user information is considered as the identification of one user. The experimental data set is a collection based on a sample of anonymized user interaction on the news feeds, collected by the Yahoo! Today module and published by Yahoo! Research Lab\footnote{http://webscope.sandbox.yahoo.com/catalog.php}.
 The dataset contains 15 days' visit events of user-news item interaction data. Besides, user's information, e.g., demographic information (age and gender), is provided for each visit event, and represented as the user identification.
 This data set has been used for evaluating contextual bandit algorithms in~\cite{li2010contextual}, \cite{li2012unbiased},\cite{tang2015personalized},\cite{zeng2016online}.
In our experiments, the visit events of the first day are utilized for selecting proper parameters of ICTR model, while 2.0 million of the remaining are used for the validation.
Each interactive record has a user ID, a news ID, the value of rating, the news pool and time stamp.

\textbf{MovieLens (10M):} Online movie recommender service aims to maximize the customer's satisfaction by recommending the proper movies to the target users according to their preference.
Specifically,
several movies out of a movie pool are first displayed to users, and then users' feedback on displayed movies is collected for improving the user satisfaction.
Thereby, the problem of movie recommendation can be regarded as an instance of bandit problem in which an arm is a movie, a pull is regarded as a movie selection, and the reward indicates the feedbacks of user's rating on movies.
In our experiments, each rating associates a user ID, a movie ID, and a time stamp.
In order to use the \emph{replayer} evaluation method, each value of a rating is processed to be 1.0 if the rating is larger than 4.0/5.0 and 0.0 otherwise.
Additionally, top-N $(N=100)$ popular movies are selected as the movie pool.

\subsubsection{Evaluation Method}
We apply the \emph{replayer} method to evaluate our proposal method on the aforementioned two datasets.
The \emph{replayer} method is first introduced in \cite{li2012unbiased}, which provides an unbiased offline evaluation via the historical logs. The main idea of \emph{replayer} is to replay each user visit to the algorithm under evaluation. If the recommended item by the testing algorithm is identical to the one in the historical log, this visit is considered as an impression of this item to the user. The ratio between the number of user clicks and the number of impressions is referred to as CTR. The work in \cite{li2012unbiased} shows that the CTR estimated by the \emph{replayer} method approaches the real CTR of the deployed online system.

\subsubsection{CTR/Rating Optimization for News/Movies Recommendation}
\begin{figure}[htp!]
\centering
  \scalebox{0.22}{
   \includegraphics{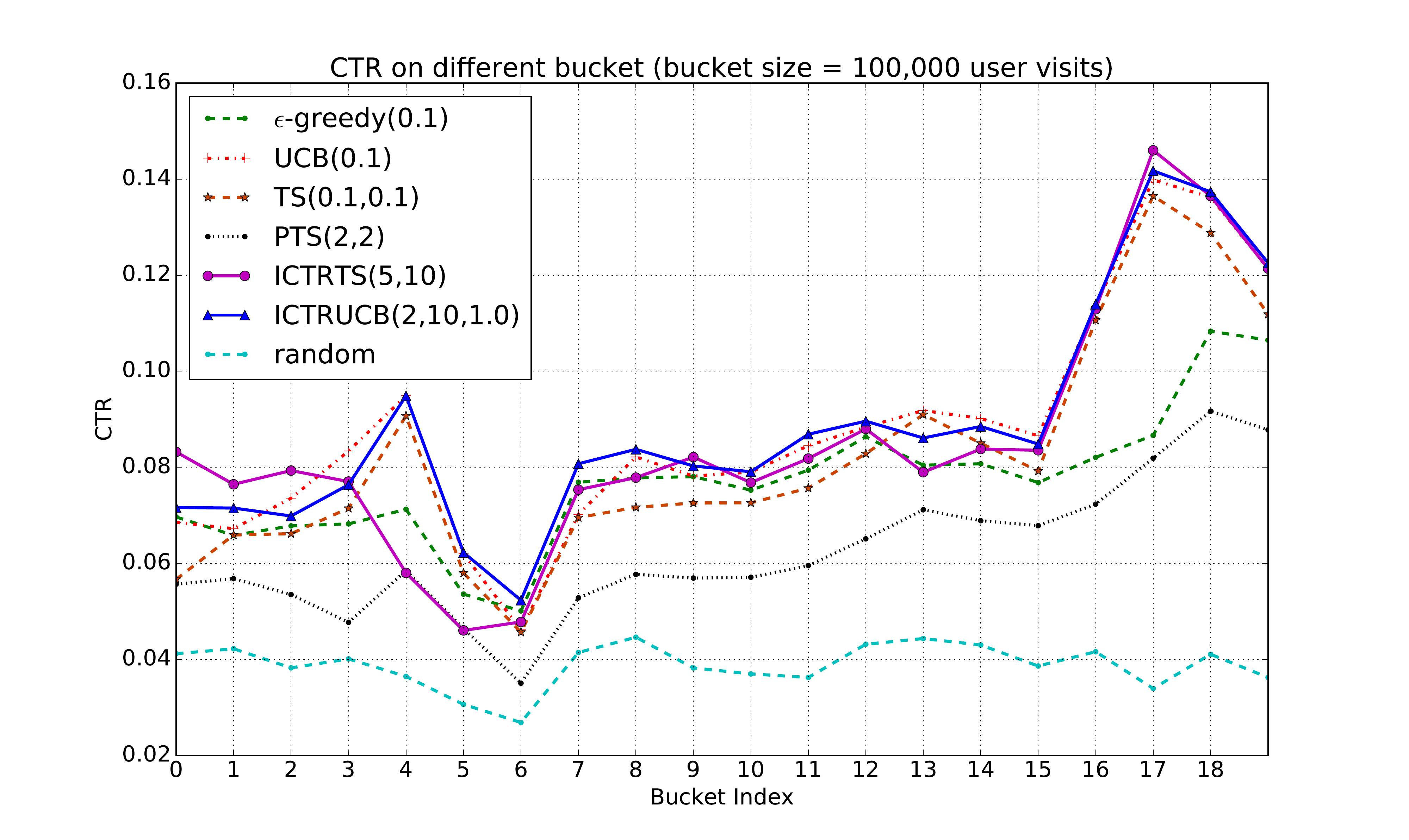}
  }
\vspace{-0.2in}
\caption{The CTR of Yahoo! Today News data is given along each time bucket. All algorithms shown here are configured with their best parameter settings.}
\label{fig:ydata-performance}
\end{figure}

\begin{figure}[htp!]
 \centering
  \scalebox{0.22}{
   \includegraphics{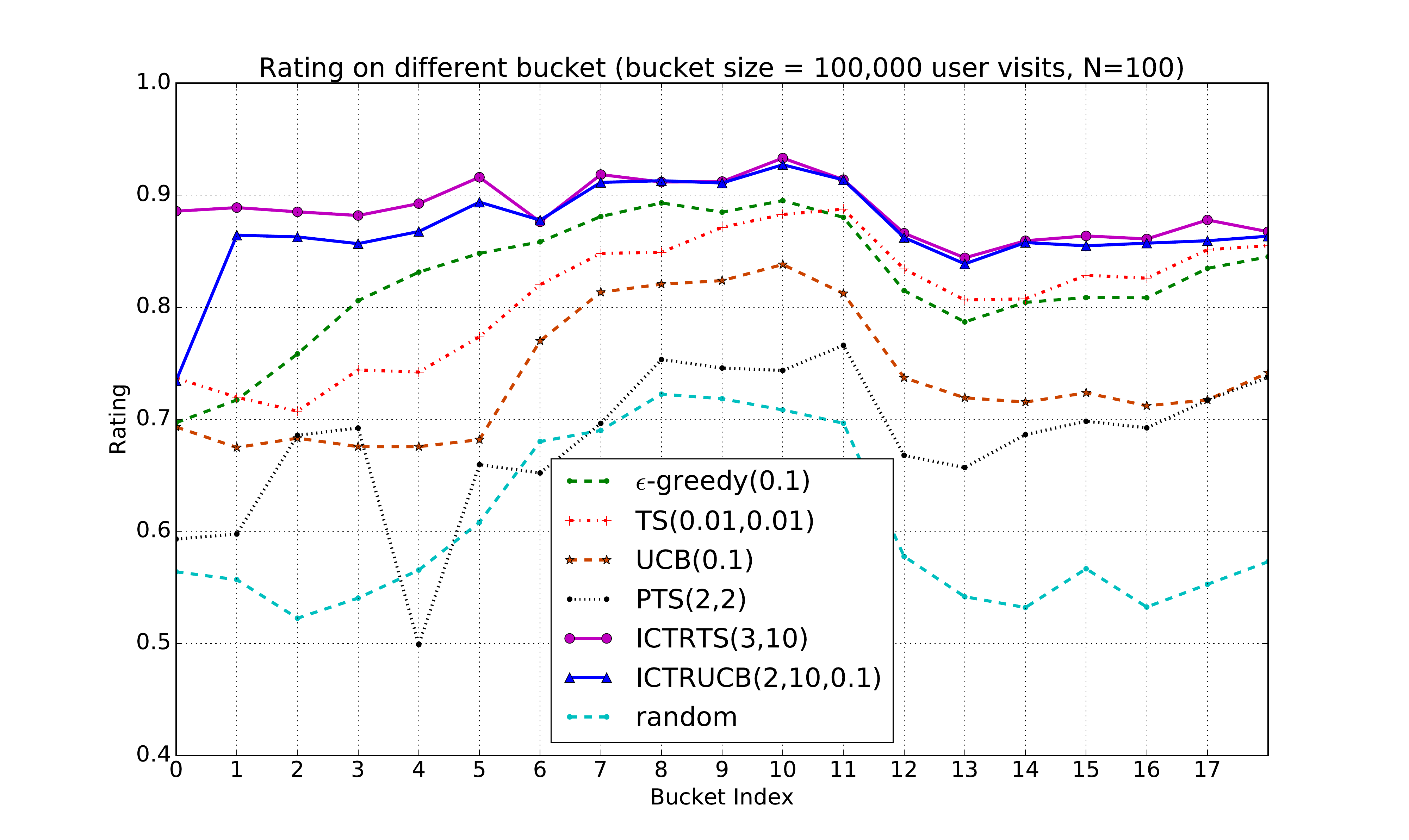}
  }
\vspace{-0.2in}
\caption{The Rating of MovieLens (10M) data is given along each time bucket. All algorithms shown here are configured with their best parameter settings.}
\label{fig:movielen-performance}
\end{figure}

In this section we first conduct the performance evaluation for each algorithm with different parameter settings. All baseline algorithms are configured with their best parameter settings provided by Table~\ref{tab:relativectr}. The setting of all algorithms with the highest reward is highlighted in bold. Our algorithm \texttt{ICTRUCB(2,10,0.1)} achieves the best performance among all algorithms on Yahoo! Today News, and performance comparisons on different algorithms along different time buckets are shown in Figure~\ref{fig:ydata-performance}. For MovieLens (10M), \texttt{ICTRTS(3,10)} outperforms all others and the corresponding performance comparisons are shown in Figure~\ref{fig:movielen-performance}.

\begin{table*}[!htb]
        \centering
        \caption{Results of Relative CTR (Rating) on two real world datasets}
        \vspace{-0.05in}
        \scalebox{0.9}{
        \begin{tabular}{llllllllll}
        \toprule
        \textbf{Algorithm} & \multicolumn{4}{c}{\textbf{Yahoo! Today News}}  &\quad\quad& \multicolumn{4}{c}{\textbf{MovieLens (10M)}}\\
        \cmidrule(r){2-5} \cmidrule(r){7-10}
           & mean  &  std  & min & max  && mean  &  std  & min & max \\
        \midrule
        \texttt{$\epsilon$-greedy(0.01)} & 0.06916 & 0.00312 & 0.06476 & 0.07166  && 0.70205 & 0.06340 & 0.60752 & 0.78934 \\
        \texttt{$\epsilon$-greedy(0.1)} & \textbf{0.07566} & 0.00079 & 0.07509 & 0.07678   && \textbf{0.82038} & 0.01437 & 0.79435 & 0.83551 \\
        \texttt{$\epsilon$-greedy(0.3)} & 0.07006 & 0.00261 & 0.06776 & 0.07372   && 0.80447 & 0.01516 & 0.77982 & 0.82458 \\
        \texttt{$\epsilon$-greedy(1.0)} & 0.03913 & 0.00051 & 0.03842 & 0.03961   && 0.60337 & 0.00380 & 0.59854 & 0.60823 \\
        \midrule
        \texttt{UCB(0.01)} & 0.05240 & 0.00942 & 0.04146 & 0.06975  && 0.62133 & 0.10001 & 0.45296 & 0.73369 \\
        \texttt{UCB(0.1)} & \textbf{0.08515} & 0.00021 & 0.08478 & 0.08544   && \textbf{0.73537} & 0.07110 & 0.66198 & 0.85632 \\
        \texttt{UCB(0.5)} & 0.05815 & 0.00059 & 0.05710 & 0.05893   && 0.71478 & 0.00294 & 0.63623 & 0.64298 \\
        \texttt{UCB(1.0)} & 0.04895 & 0.00036 & 0.04831 & 0.04932   && 0.63909 & 0.00278 & 0.60324 & 0.61296 \\
        \midrule
        \texttt{TS(0.01,0.01)} & 0.07853 & 0.00058 & 0.07759 & 0.07921   && \textbf{0.83585} & 0.00397 & 0.82927 & 0.84177\\
        \texttt{TS(0.1,0.1)} & \textbf{0.07941} & 0.00040 & 0.07869 & 0.07988   && 0.83267 & 0.00625 & 0.82242 & 0.84001\\
        \texttt{TS(0.5,0.5)}  & 0.07914 & 0.00106 & 0.07747 & 0.08041   && 0.82988 & 0.00833 & 0.81887 & 0.84114\\
        \texttt{TS(1.0,1.0)} & 0.07937 & 0.00079 & 0.07788 & 0.08044   && 0.83493 & 0.00798 & 0.82383 & 0.84477\\
        \midrule
        \texttt{PTS(2,2)} & \textbf{0.06069} & 0.00575 & 0.05075 & 0.06470   && \textbf{0.70484} & 0.03062 & 0.64792 & 0.74610 \\
        \texttt{PTS(2,10)} & 0.05699 & 0.00410 & 0.05130 & 0.06208   && 0.65046 & 0.01124 & 0.63586 & 0.66977 \\
        \texttt{PTS(5,10)} & 0.05778 & 0.00275 & 0.05589 & 0.06251   && 0.63777 & 0.00811 & 0.62971 & 0.65181 \\
        \texttt{PTS(5,20)} & 0.05726 & 0.00438 & 0.05096 & 0.06321   && 0.62289 & 0.00714 & 0.61250 & 0.63567 \\
        \texttt{PTS(10,20)} & 0.05490 & 0.00271 & 0.05179 & 0.05839   && 0.61819 & 0.01044 & 0.60662 & 0.63818 \\
        \midrule
        \texttt{ICTRTS(2,5)} & 0.06888 & 0.00483 & 0.06369 & 0.07671   && 0.70386 & 0.15772 & 0.48652 & 0.85596 \\
        \texttt{ICTRTS(2,10)} & 0.06712 & 0.01873 & 0.03731 & 0.08487   && 0.56643 & 0.10242 & 0.42974 & 0.67630 \\
        \texttt{ICTRTS(3,10)} & 0.06953 & 0.00783 & 0.05857 & 0.07804   && \textbf{0.88512} & 0.00052 & 0.88438 & 0.88553 \\
        \texttt{ICTRTS(5,10)} & \textbf{0.08321} & 0.08236 & 0.08492 & 0.06292   && 0.55748 & 0.14168 & 0.38715 & 0.73404 \\
        \texttt{ICTRTS(7,10)} & 0.05066 & 0.00885 & 0.04229 & 0.06423   && 0.517826 & 0.07120 & 0.42297 & 0.59454 \\
        \texttt{ICTRTS(7,20)} & 0.04925 & 0.00223 & 0.04672 & 0.05285   && 0.61414 & 0.12186 & 0.44685 & 0.73365 \\
        \midrule
        \texttt{ICTRUCB(2,10,0.01)} & 0.06673 & 0.01233 & 0.04588 & 0.08112   && 0.44650 & 0.06689 & 0.38678 & 0.53991 \\
        \texttt{ICTRUCB(2,10,1.0)} & \textbf{0.08597} & 0.00056 & 0.08521 & 0.08675   && \textbf{0.86411} & 0.01528 & 0.85059 & 0.88547 \\
        \texttt{ICTRUCB(3,10,0.05)} & 0.07250 & 0.00426 & 0.06799 & 0.07694   && 0.54757 & 0.13265 & 0.43665 & 0.73407 \\
        \texttt{ICTRUCB(3,10,1.0)} & 0.08196 & 0.00296 & 0.07766 & 0.08530   && 0.57805 & 0.08716 & 0.46453 & 0.67641 \\
        \texttt{ICTRUCB(5,10,0.01)} & 0.07009 & 0.00722 & 0.06411 & 0.08244   && 0.62282 & 0.02572 & 0.59322 & 0.65594 \\
        \texttt{ICTRUCB(5,10,1.0)} & 0.08329 & 0.00140 & 0.08098 & 0.08481   && 0.80038 & 0.24095 & 0.9625 & 0.88554 \\
        \bottomrule
        \end{tabular}
        }
        \label{tab:relativectr}
\end{table*}

\subsubsection{A Case Study: Movie Topic Distribution Analysis on MovieLens (10M)}
We conduct an experiment to demonstrate that our model can effectively capture the dependency between items, i.e., finding the latent topics among movies and clustering similar movies together. In this experiment, top-N $(N=8)$ popular movies are selected and topic number $(K=2)$ is set as config to our model. After millions of training iterations, the learned latent movie feature vectors will represent each movie's topic distribution over the two latent topics, in which the $i$-th dimension of the feature vector encodes the probability that the movie belongs to the $i$-th movie topic cluster.
We separately choose four movies with the highest value of the first element and the second element of these latent feature vectors, and list their IDs, names, and movie types in Table~\ref{tab:moviecluster}, which clearly proves our assumption that the model is able to capture the dependency between items and cluster similar movies together.
\begin{table*}[!htb]
        \vspace{-0.1in}
        \centering
        \caption{Movie Topic Distribution of MovieLen (10M)}
        \scalebox{0.9}{
        \begin{tabular}{ccccccc}
        \toprule
        \multicolumn{3}{c}{\textbf{Topic Cluster I}}  &\quad\quad  & \multicolumn{3}{c}{\textbf{Topic Cluster II}} \\
        \cmidrule(r){1-3} \cmidrule(r){5-7}
         MovieId  &  MovieName  & MovieType && MovieId  &  MovieName  & MovieType  \\
        \midrule
         32 & 12 Monkeys & Sci-Fi,Thriller    && 344 & Pet Detective & Comedy \\
         \midrule
         50 & Usual Suspects & Crime,Mystery,Thriller    && 588 & Aladdin & Children,Animation,Comedy \\
         \midrule
         590 & Dances with wolves & Adventure,Drama,Western    && 595 & Beauty and the Beast & Animation,Children,Musical\\
         \midrule
         592 & Batman & Action,Crime,Sci-Fi,Thriller    && 2857 & Yellow Submarine & Adventure,Animation,Comedy,Musical \\
        \bottomrule
        \end{tabular}
        }
        \label{tab:moviecluster}
\end{table*}
\subsubsection{Time Cost}
\begin{figure}[htp!]
\centering
\begin{subfigure}{0.23\textwidth}
  \scalebox{0.11}{
   \includegraphics{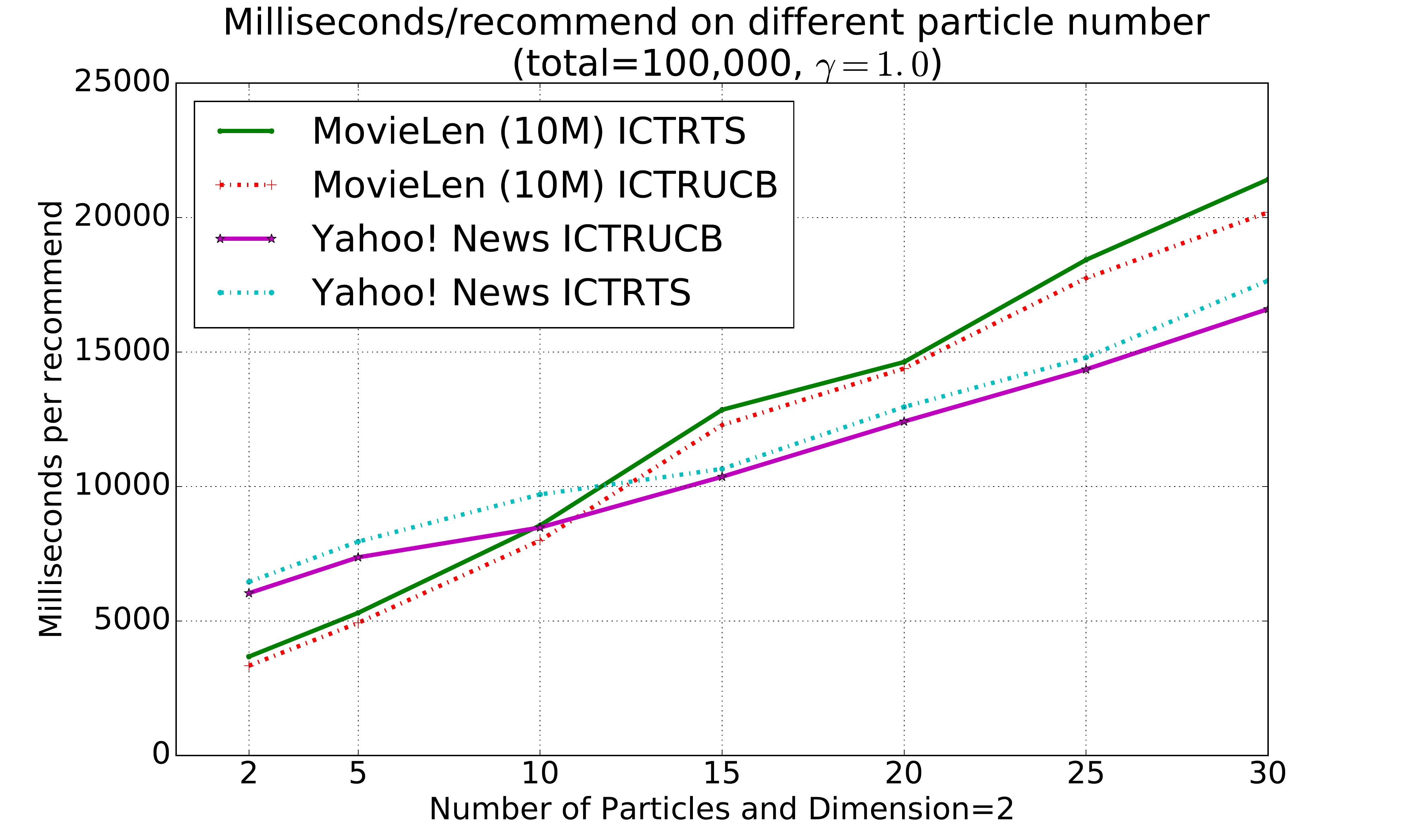}
  }
\label{fig:timecostparticle}
\end{subfigure}
\begin{subfigure}{0.23\textwidth}
  \scalebox{0.11}{
   \includegraphics{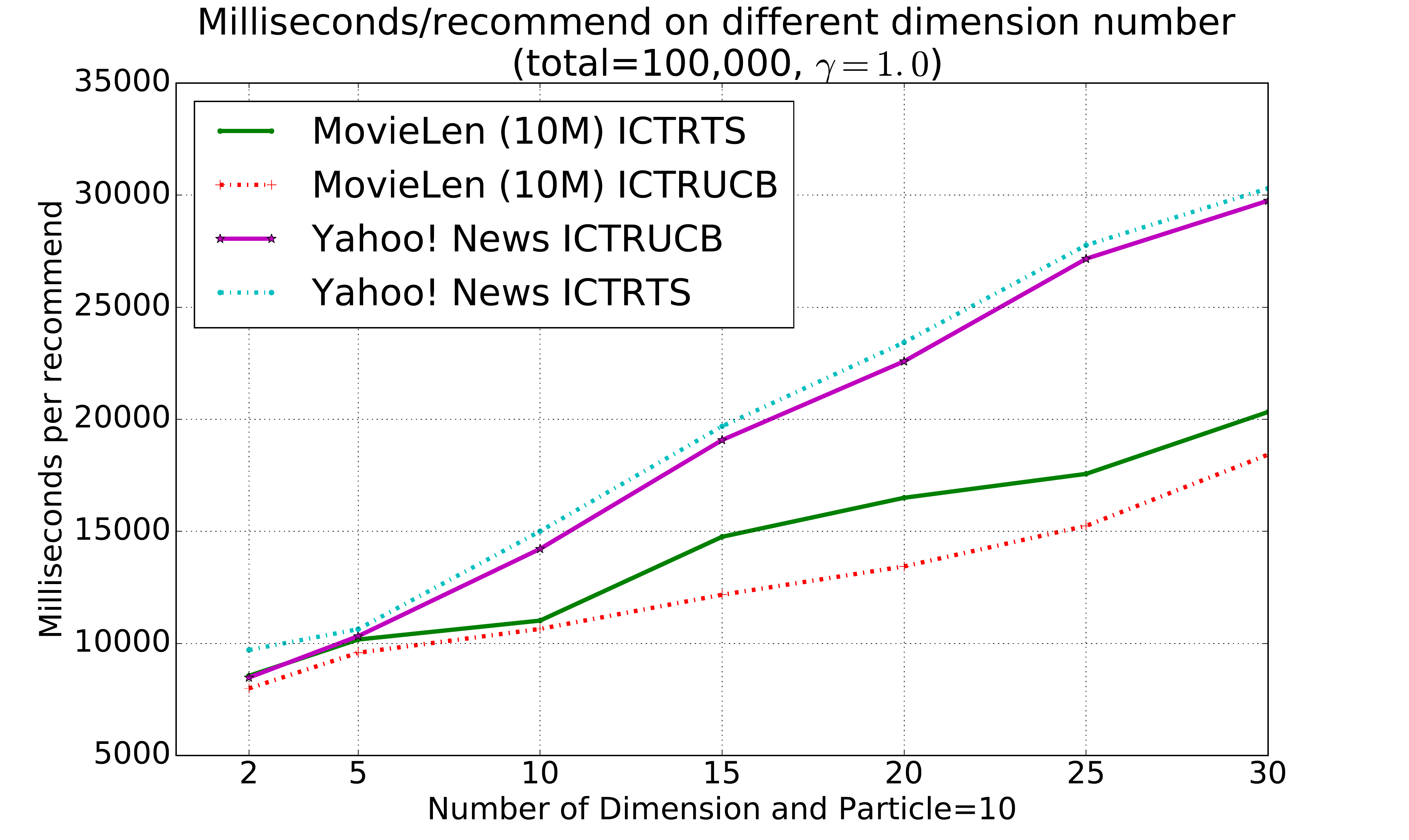}
  }
\label{fig:timecostdimension}
\end{subfigure}
\vspace{-0.2in}
\caption{Time cost are given with respect to different number of particles and latent feature vector dimensions for both two datasets.}
\label{fig:timecost}
\vspace{-0.2in}
\end{figure}

The time cost for \texttt{ICTRTS} and \texttt{ICTRUCB} on both two data sets is displayed in Figure~\ref{fig:timecost}. It shows that the time costs are increased linearly with the number of particles and dimensions of latent feature vector. MovieLens (10M) costs  much more time than Yahoo! Today New due to a larger amount of items and users. In general, \texttt{ICTRUCB} is faster than \texttt{ICTRTS} since \texttt{ICTRTS} highly depends on the sampling process.

\section{Conclusion and Future Work} \label{sec:conclusion}

In this paper, we propose an interactive collaborative topic regression model that adopts a generative process based on topic modeling to explicitly formulate the arm dependencies as the clusters on arms, where dependent arms are assumed to be generated from the same cluster. Every time an arm is pulled, the feedback is not only used for inferring the involved user and item latent vectors, but also employed to update the latent parameters with respect to the arm's cluster. The latent cluster parameters further help with the reward prediction for other arms in the same cluster. We conduct empirical studies on two real-world applications, including movie and news recommendation, and the experimental results demonstrate the effectiveness of our proposed approach.

\noindent \ \ \ \ Individual preferences on news and movies usually evolve over time. One possible research direction is to extend our model considering the time-varying property in user preferences for better online personal recommendation~\cite{zeng2016online}. In this paper, we formulate the arm dependencis as soft clustering. Another possible research direction is to explicitly eplore the hirerarchial structure among arms~\cite{zeng2017knowledge} (i.e. items in specific domain areas), which can be manually or dynamically constructed using domain knowledge~\cite{wang2017constructing}.

\bibliographystyle{ACM-Reference-Format}
\bibliography{sigproc}

\end{document}